\documentclass[twocolumn,english,aps,prb,showpacs,byrevtex,amsmath,amssymb,superscriptaddress]{revtex4-1}
\usepackage{fancyhdr}
\usepackage{amsmath}
\usepackage{subfigure}
\usepackage{epsfig}
\usepackage{amssymb}
\usepackage{units} 
\usepackage{pifont}
\usepackage{textcomp}
\usepackage{gensymb}
\usepackage{pifont}
\usepackage{enumerate}
\usepackage{color}
\usepackage{float}
\usepackage{bm}
\usepackage{ulem}
\begin{document}

\title{Double-helix magnetic order in CrAs with Pnma space group}

\author{Giuseppe Cuono}
\email{gcuono@magtop.ifpan.edu.pl}

\affiliation{International Research Centre Magtop, Institute of Physics, Polish Academy of Sciences,
Aleja Lotnik\'ow 32/46, PL-02668 Warsaw, Poland}

\author{Alfonso Romano}
\affiliation{Dipartimento di Fisica "E.R. Caianiello", Universit\`a degli Studi di Salerno, I-84084 Fisciano
(SA), Italy}
\affiliation{Consiglio Nazionale delle Ricerche CNR-SPIN, UOS Salerno, I-84084 Fisciano (Salerno),
Italy}

\author{Canio Noce}
\affiliation{Dipartimento di Fisica "E.R. Caianiello", Universit\`a degli Studi di Salerno, I-84084 Fisciano
(SA), Italy}
\affiliation{Consiglio Nazionale delle Ricerche CNR-SPIN, UOS Salerno, I-84084 Fisciano (Salerno),
Italy}

\author{Carmine Autieri}
\email{autieri@magtop.ifpan.edu.pl}
\affiliation{International Research Centre Magtop, Institute of Physics, Polish Academy of Sciences,
Aleja Lotnik\'ow 32/46, PL-02668 Warsaw, Poland}
\affiliation{Consiglio Nazionale delle Ricerche CNR-SPIN, UOS Salerno, I-84084 Fisciano (Salerno),
Italy}

\date{\today}
\begin{abstract}
		
The transition metal pnictide CrAs exhibits superconductivity in the vicinity of an helimagnetic phase, where it has been found that the propagation vector is parallel to the $c$ axis and the magnetic moments lie in the $ab$ plane. Here we use {\it ab initio} calculations to study the magnetic interactions in the material. Mapping onto an Heisenberg Hamiltonian, we calculate the magnetic exchanges by using LDA+U calculations and we unveil the origin of the magnetic frustration. Finally, we reproduce the double helix magnetic order with a propagation vector $\bm Q=(0,0,0.456)$ and we obtain the magnetic transition temperature $T_N$ through Monte-Carlo simulations of the specific heat.
Due to the limitations of the use of the Heisenberg Hamiltonian for itinerant magnetic systems, the theoretical $T_N$ underestimated the experimental value of the pure CrAs. However, our results are in good agreement with those found for the alloy CrAs$_{0.5}$Sb$_{0.5}$ belonging to the same space group, showing that our result can describe this material class.

\end{abstract}

\maketitle

\section{Introduction}

In recent years an increasing interest has been devoted to transition metal pnictides \cite{Chen19,Noce20}. These materials show unconventional properties determined by the existence of a superconducting phase which develops in proximity of a magnetic one that can be suppressed through the application of a physical or a chemical pressure \cite{Wu14,Wang16,Wu10,Kotegawa14}. 
Some materials of this family, such as CrAs \cite{Wu10,Wu14,Kotegawa14,Autieri17,Autieri17b,Autieri18,Cuono19c,Nigro19}, MnP \cite{Cheng15} and WP \cite{Liu19,Cuono19,Nigro20}, belong to the space group Pnma, while another family is represented by the compounds  A$_2$Cr$_3$As$_3$ \cite{Bao15,Cuono18,Cuono19b,Cuono21a,Cuono21b,Cuono21c}, with the element A being Na \cite{Mu18}, K \cite{Bao15}, Rb \cite{Tang15} or Cs \cite{Tang15b}, which have in common with the first group the possible presence of unconventional superconductivity. In particular, in the process of chemical cation deintercalation applied to K$_2$Cr$_3$As$_3$ to obtain the magnetic parent compound KCr$_3$As$_3$, an intermediate state between the 233 and the 133 phases has been observed \cite{Zhang21,Galluzzi21}, possibly allowing to tune the interplay between magnetic and superconducting properties \cite{Autieri12}.    
%
%
%
Generally speaking, we can say that though these systems are being thoroughly investigated to date, they still need a careful analysis to better understand the interplay between structural, electronic, magnetic and superconducting properties and their connections to topological features.

The material class with Pnma phase (space group number 62) presents nonsymmorphic symmetries that are responsible for additional degeneracies of the bands \cite{Cuono19,Cuono19b}. This affects the shape of the Fermi surface, as well as the magnetic and the superconducting properties, giving rise, together with the presence of strong magnetic fluctuations, to a quasilinear magnetoresistance (MR) \cite{Niu17} near the magnetic instability.
There are very few cases of non-compensated materials with topological band structure showing large MR \cite{Campbell21}, while a non-saturating extremely large MR due to the presence of compensated Fermi pockets is more common in nature\cite{Wadge21}.
Regarding CrAs, pressure-induced superconductivity was found at $T_c\sim\,$2 K and $P_c\sim\,$8 kbar in proximity of an helimagnetic (HM) phase \cite{Wu14,Kotegawa14}. 
In the non-collinear HM one, it has been found by neutron diffraction measurements \cite{Shen16,Keller15} that the propagation vector is parallel to the c axis and the magnetic moments lie in the $ab$ plane. The transition to the HM phase occurs at $T_N\sim\,$275 K.
Four spirals propagating along the $c$ axis with a well defined angle between them form the magnetic structure of the compound, and the magnetic moment decreases with pressure from 1.7 $\mu_B$ to 0.4 $\mu_B$ at a pressure of 0.7 GPa, when superconductivity starts to appear and the magnetic order is suppressed \cite{Keller15}. This decrease of the magnetic moment is accompanied by a spin reorientation, from a configuration where the magnetic moments lie in the $ab$ plane to one where they lie in the  $ac$ plane \cite{Shen16}. We recall that in these compound magnetism is due to the $d$ orbitals of the Cr atoms at the Fermi level\cite{Autieri17,Autieri17b,Autieri18}.

In CrAs the magnetic, structural and electronic properties are intimately related, as made evident by the fact that the magnetic transition is accompanied by discontinuous changes of the lattice parameters \cite{Wu14}. These effects have been analyzed via {\it ab initio} studies \cite{Autieri17} and model Hamiltonian investigations \cite{Autieri17b,Autieri18,Cuono19c} which allowed to conclude that CrAs is a metal with moderate electronic correlations.
Indeed, the band structure is well described within the Density Functional Theory (DFT) \cite{Autieri17} and a relatively small value of the Coulomb repulsion allows to obtain for the magnetic moment a value in agreement with the experimental one \cite{Autieri17b}.

The non-collinear magnetic states can originate from the Dzyaloshinskii-Moriya interaction or from magnetic frustrations.
Several examples of the latter are present in transition metal compounds
\cite{Ming,Ivanov,Sliwa1,Sliwa2}; depending on the symmetries and the degree of frustration, the system can show spin-spiral, double helix, spin glass or a collinear magnetic phase in the case of low frustration.
In particular, many transition metal monopicnitides show double helix magnetic order\cite{Wang16}, with this kind of order having been recently found also in CrAs. From inelastic neutron scattering data on polycrystalline samples, an oversimplified model for the exchange couplings was derived in order to demonstrate the magnetic frustration origin of the double spin helix.\cite{Matsuda18}
However, a study within DFT of the magnetic frustration and of the double helix order is still missing in the literature.

In this paper, we improve the theoretical description of the double helix configuration in CrAs presented in Ref.~\onlinecite{Matsuda18}, calculating the magnetic couplings via the {\it ab initio} DFT+U method mapped onto an Heisenberg model. We unveil the origin of the magnetic frustration that originates the double helix, reliably simulating the related helical magnetic order. The validity of our approach has been tested evaluating the specific heat and the transition temperature $T_N$ through the Monte-Carlo method. 
We show that, although the system is metallic, a simple Heisenberg model allows to capture fundamental properties such as the frustration between magnetic exchanges of different nature and the double helical structure.

The paper is organized as follows: in the next section we report the computational details, in the third section the results and in the last section the conclusions.

\section{Computational details}

We have performed Density Functional Theory calculations by using the VASP package \cite{Kresse93,Kresse94,Kresse96a,Kresse96b}. In this approach, the core and the valence electrons have been treated within the Projector Augmented Wave (PAW) method \cite{Kresse99} and with a cutoff of 400 eV for the plane wave basis. All the calculations have been performed
using a 12$\times$16$\times$10 $k$-point grid. For the treatment of the exchange correlation, the Local Density Approximation (LDA) and the Perdew-Zunger parametrization \cite{Perdew81} of the Ceperley-Alder \cite{Ceperley80} data have been considered since they usually give better results for itinerant magnetic systems\cite{Wyso21}.
In DFT, the electronic interactions are already present as the sum of the Hartree term and the exchange correlation that include all the correlations and spin interactions. When weakly correlated compounds are studied, it is not necessary to consider additional interactions. When moderately and strongly correlated electron systems are instead considered, an additional Coulomb repulsion $U$ is added to the energy functional \cite{Liechtenstein95}. This leads to an increase of the magnetic moment with respect to what one finds with the simple LDA, so that in the case of moderate correlations the values of $U$ must be carefully chosen in order to avoid an overestimation of the magnetic moment.
For the system analyzed here, we have found that the best agreement with the experiments is obtained for $U=1\,$eV. 

In the evaluation of the magnetic exchanges of CrAs, we found that not all the magnetic configurations converge in the case of collinear magnetism. However, the inclusion of the spin-orbit coupling favored the convergence, allowing to determine a set of magnetic configurations sufficient to obtain the relevant exchange couplings needed to describe the double helix magnetic order.

\section{Results}

In this section, we first present the calculation of the magnetic couplings and then we analyze the double-helix magnetic order in CrAs, evaluating in particular the specific heat by use of MonteCarlo simulations.

\subsection{Electronic properties and magnetic exchanges}

\begin{figure}
	\centering
	\includegraphics[width=8.2cm]{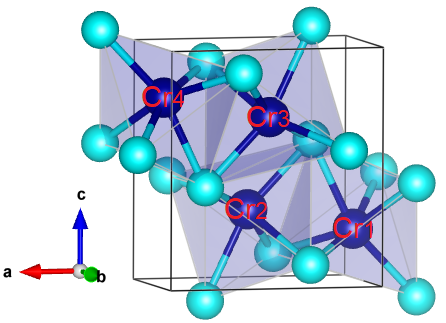}
	\caption{Crystal structure of CrAs in the MnP-type phase. Cr and As atoms are shown as blue and cyan spheres, respectively. The CrAs$_6$ face-shared  octahedra are transparent. The solid black lines represent the orthorhombic primitive unit cell.}
	\label{structure}
\end{figure}

CrAs belongs to the family of transition-metal pnictides sharing the general formula AB (A = transition metal, B = P, As, Sb). It exhibits either a hexagonal NiAs-type (B81) structure or an orthorhombic MnP-type (B31) structure. In the latter case, the unit-cell lattice parameters are $a$= 5.649 {\AA}, $b$ =  3.463 {\AA} and $c$ = 6.2084 {\AA}. The Cr atoms are situated in the center of the CrAs$_6$ octahedra, surrounded by six nearest-neighbour arsenic atoms, as shown in Fig.~\ref{structure}; four of the six Cr–As bonds are inequivalent due to the high anisotropy exhibited by this class of compounds.

\begin{figure}
	\centering
	\includegraphics[width=9.2cm]{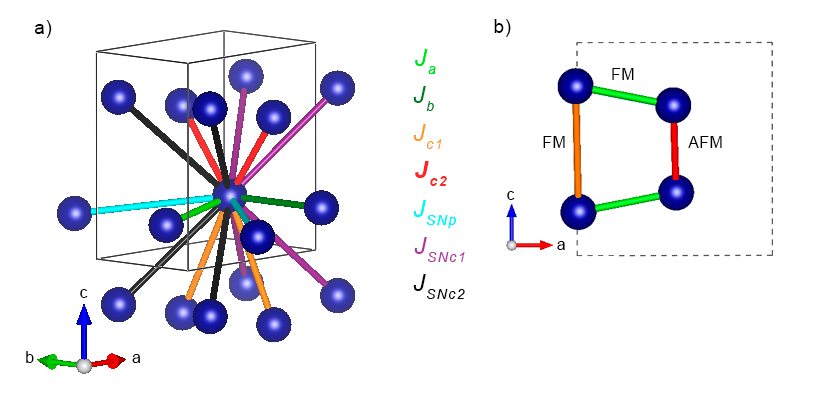}
	\caption{Magnetic couplings considered in our calculation. In panel a) we show the 3D sight (the colors of the $J$ denominations coincide with the colors of the bonds), while in panel b) that in the $ac$ plane where we show the characteristic Cr-Cr distances. With FM we indicate the ferromagnetic coupling, while with AFM the antiferromagnetic one.}
	\label{fig:couplings}
\end{figure}

The band structure of CrAs obtained in LDA \cite{Autieri17} is characterized by
a bandwidth of 3.5 eV. We can compare this value with the one of the t$_{2g}$ bands in strongly correlated Cr-based systems, equal to 2.5 eV\cite{Autieri14K,Asa1,Asa2}, as well as with the one of the uncorrelated elemental Cr which is around 5.0 eV\cite{Jain13,Asa3}. This allows to conclude that CrAs is a moderately correlated system. We have thus performed {\it ab initio} LDA+U calculations scanning a range of low values of $U$, eventually obtaining the best agreement with the experimental data on the exchange couplings for $U=1\,$eV. 
%
We have used the Hamiltonian \cite{Matsuda18}
\begin{equation}
\label{eqn:exchange}
H=\sum_{i,j}J_{ij}\,\boldsymbol{S}_{i}\cdot\boldsymbol{S}_{j} \; ,
\end{equation} 
considering the first-neighbours exchange couplings, denoted as $J_{a}$, $J_{b}$, $J_{c1}$ and $J_{c2}$, the in-plane second-neighbour coupling $J_{SNp}$, and the second-neighbour couplings along $c$ $J_{SNc1}$ and $J_{SNc2}$, as shown in Fig.~\ref{fig:couplings}a.

The collinear magnetic phase closer to the double-helix ground state was investigated before and found to be of G-type \cite{Autieri17b,Autieri18}. 
Here we have considered the ferromagnetic phase together with three other magnetic phases with zero net magnetic moment which are usually considered in G-type, C-type and A-type Pnma perovskites.
With reference to the eight-atom unit cell, the four equations for the above mentioned four magnetic configurations are

\small
\begin{eqnarray*}
E_{FM} & = & E_1 + 4 J_a + 4 J_{c1} + 4 J_{c2} + 8 J_{SNp} + 8 J_{SNc1} + 8 J_{SNc2} \\
E_{G} & = & E_1 - 4 J_a - 4 J_{c1} - 4 J_{c2} - 8 J_{SNp} + 8 J_{SNc1} + 8 J_{SNc2} \\
E_{C} & = & E_1 - 4 J_a + 4 J_{c1} + 4 J_{c2} - 8 J_{SNp} - 8 J_{SNc1} - 8 J_{SNc2} \\
E_{A} & = & E_1 + 4 J_a - 4 J_{c1} - 4 J_{c2} + 8 J_{SNp} - 8 J_{SNc1} - 8 J_{SNc2}
\end{eqnarray*}
\normalsize

\noindent where $E_1$ is a trivial constant energy and the other quantities are the magnetic exchanges shown in Fig.~\ref{fig:couplings}a.
In these four equations, we have seven unknowns quantities not including $J_b$; moreover, we can estimate just the quantity $J_{c1}$+$J_{c2}$ but not the single couplings $J_{c1}$ and $J_{c2}$. 


\begin{table} 
\begin{center}
    \begin{tabular}{| l | l | l | l | l | l | l | l |}
    \hline
    \multicolumn{1}{ |c| }{} & \multicolumn{4}{ |c| }{ First neighbours} & \multicolumn{3}{ |c| }{Second neighbours} \\
\hline
Distance & 2.857 & 3.588 & 4.042 & 3.090 & 4.586 & 4.478 & 4.574   \\ \hline
 & $J_a$ & $J_b$ & $J_{c1}$ & $J_{c2}$ & $J_{SNp}$ & $J_{SNc1}$ & $J_{SNc2}$   \\ \hline
    LDA + U & -19.4 & -25.6  & -21.9 &  58.6 & 5.6 & -8.2 & -8.2  \\ \hline
    Ref.~\onlinecite{Matsuda18} & 9.27 &  & -4.82 & 65.0 &  &  &   \\
    \hline
    \end{tabular}
\end{center}
\caption{In the first line we indicate the distances in {\AA} between neighboring , in the third and the fourth line we report the values of $J$ obtained with our calculations in LDA+U approximation and the values of Ref.~\onlinecite{Matsuda18} respectively. $J_a$, $J_b$, $J_{c1}$ and $J_{c2}$ refer to the first neighbours along $a$, $b$ and $c$ axes; $J_{SNp}$ to the in-plane second neighbour;  $J_{SNc1}$ and $J_{SNc2}$ to the second neighbours along $c$. The latter are two independent parameters which in our approximation are assumed to be equal. Since they are not the dominant exchange couplings, this approximation will not influence the main results. The values of $J$ are expressed in meV, while the one of the Coulomb repulsion is in eV. \label{tabJ}}
\end{table}

The DFT calculations do not converge easily once we go away from the G-type magnetic configuration or away from the zero magnetic moment ones. Therefore, to estimate $J_b$ we have doubled the unit cell along the $b$ axis, but we have taken into account only configurations with zero net magnetic moment. In particular, we have considered two unit cells with G-type magnetic order, one coupled ferromagnetically and the other one antiferromagnetically, the related energies being denoted as $E_{GFMb}$ and  $E_{GAFMb}$. We have repeated the same calculation for the A-type configuration coupled antiferromagnetically along the $b$-axis (the energy is $E_{AAFMb}$), in this way obtaining the following equations:

\begin{eqnarray*}
E_{GFMb} & = & 2 E_2  - 8 J_a + 8 J_b - 8 J_{c1} +  \\
& & -  8 J_{c2} - 16 J_{SNp} + 16 J_{SNc1} + 16 J_{SNc2}  \\
E_{GAFMb} & = & 2 E_2  - 8 J_a - 8 J_b + 16 J_{SNp} \\
E_{AAFMb} & =  & 2 E_2 + 8 J_a - 8 J_b - 16 J_{SNp} 
\end{eqnarray*}

The quantity $E_2$ is a trivial constant energy too, but its value is different from $E_1$. To estimate $J_{c1}$+$J_{c2}$, we have doubled the unit cell along the $c$-axis considering two unit cells with G-type magnetic order, again coupled ferromagnetically and antiferromagnetically, respectively. We define their energies as $E_{GFMc}$ and $E_{GAFMc}$, and we have the following equations:
\begin{eqnarray*}
E_{GAFMc} & = & 2 E_3 + 8 J_{c1} - 8 J_{c2} \\
E_{GFMc} & = & 2 E_3 - 8 J_{c1} + 8 J_{c2} 
\end{eqnarray*}

The quantity $E_3$ is again a trivial constant energy, different from $E_1$ and $E_2$.
Solving the previous equations, we have mapped the {\it ab initio} results onto a Heisenberg model\cite{Xiang13}, reporting the values of the magnetic exchanges in Table \ref{tabJ}. We find that $J_{c2}$ is the largest exchange coupling and it is ferromagnetic, while all the other nearest-neighbour ones are antiferromagnetic. Smaller values are found for the second-nearest neighbour couplings, as expected, with the in-plane and the out-of-plane ones which are ferromagnetic and antiferromagnetic, respectively. We point out that these results come from an energy minimization procedure and thus cannot be directly compared with the ones reported in Ref.~\onlinecite{Matsuda18} where the exchange couplings were rather treated as fitting parameters. We also notice that though this kind of mapping has limitations for metals, the large $T_{N}$ value of CrAs should ensure that the obtained results are correct.

\begin{figure}[t!]
	\centering
	\includegraphics[width=6.2cm]{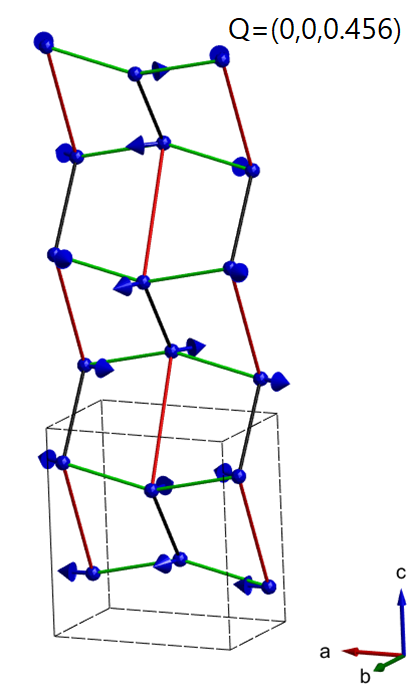}
	\caption{Double helix structure of CrAs for $\bm Q=(0,0,0.456)$ obtained with our data for the exchange couplings. The blue arrows represent the
	spin of the Cr atoms. The As atoms are not shown. The grey dashed lines represent the primitive unit cell.}
	\label{fig:helical}
\end{figure}

\begin{figure}[]
	\centering
	\includegraphics[width=\columnwidth,  angle=0]{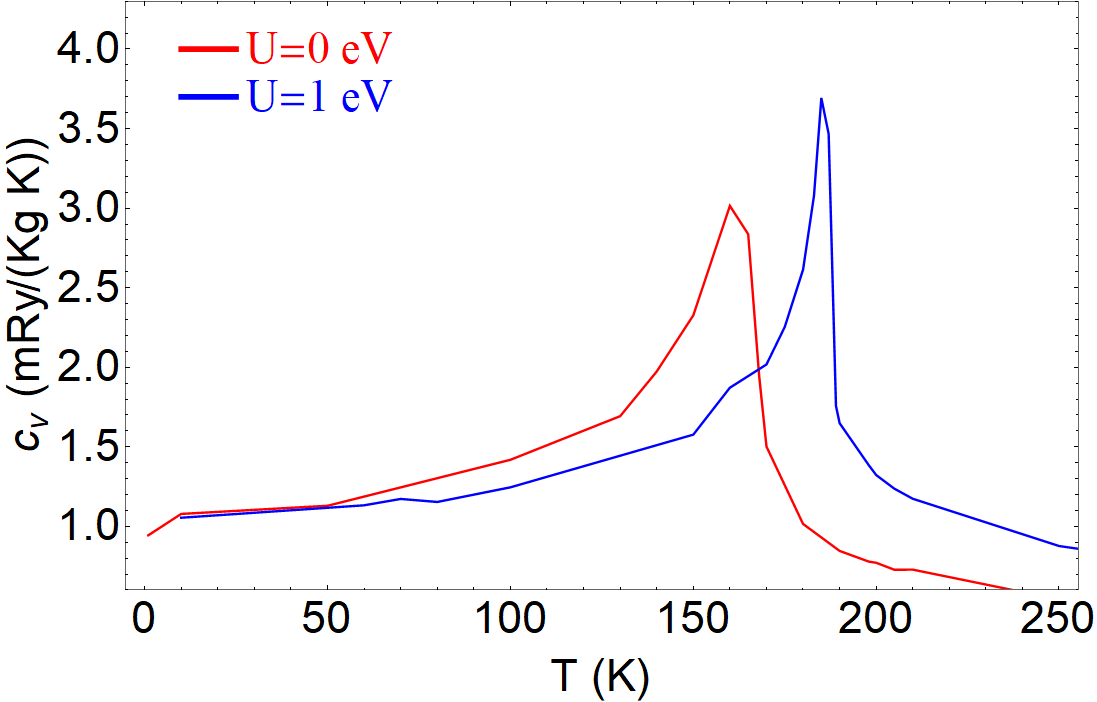}
	\caption{The red line shows the temperature dependence of the specific heat of CrAs evaluated from the magnetic exchanges calculated in LDA and the blue line shows the specific heat calculated in LDA+U approximation for $U=1\,$eV. For the curve at U=0 eV the critical temperature is $T_N$=160 K, while for the one at U=1 eV, $T_N$=185 K.}
	\label{fig:specificheat}
\end{figure}

\subsection{Double helix magnetic order}

Using the UppASD (Uppsala Atomistic Spin Dynamics) package \cite{Skubic08}, we have performed Monte Carlo simulations to evaluate the specific heat, considering a 16$\times$16$\times$16 supercell and referring to the Heisenberg-like Hamiltonian (\ref{eqn:exchange}). In CrAs there are two different characteristic Cr–Cr distances along the $c$-axis, indicated as Cr$_{2}$-Cr$_{3}$ and Cr$_{1}$-Cr$_{4}$ in Fig.~\ref{structure} and equal to 3.090 {\AA} and 4.042 {\AA}, respectively.
In the former case, i.e. in the case of shorter distance, the magnetic coupling is strongly antiferromagnetic, while in the latter one it is weakly ferromagnetic, as shown 
in Fig.~\ref{fig:couplings}b.

From Fig.~\ref{fig:couplings}b we observe that we cannot minimize the energy on all the bonds of the quadrilateral. This leads to a magnetic frustration that is responsible for the spiral configuration.
In order to reproduce the experimentally detected wave vector $\bm Q=(0, 0, 0.356)$, the stability values of the exchange coupling leading to the double helix magnetic ground state were already determined by Matsuda et al. \cite{Matsuda18} as $J_{c2}/J_{a}$=7.1 and $J_{c1}/J_{a}$=$-$0.52, irrespective of the value and the sign of $J_{b}$. 
The magnetic couplings obtained are reported in Table \ref{tabJ} together with the values obtained in Ref.~\onlinecite{Matsuda18}.
The results are different because obtained with two different methods and both methods have limitations. We used {\it ab initio} calculations and therefore we do not have a perfect agreement with the experiment, while Matsuda et al. \cite{Matsuda18} made a fit of the experimental result with a 2D model neglecting $J_{b}$ and the second neighbours.
The experimental propagating wave vector in reciprocal lattice unit is ${\bm Q}=(0, 0, 0.356)$ \cite{Matsuda18}, while the wave vector obtained with the LDA+U values of the exchange couplings is $\bm Q=(0,0,0.456)$. This value of $\bm Q$ is close to that reported by Wang et al. \cite{Wang16} for CrAs$_{x}$Sb$_{1-x}$.
The corresponding double helix structure has been calculated by means of the algorithm of Ref.~\onlinecite{Toth15} and is reported in Fig.~\ref{fig:helical}.

From the LDA+U values obtained for the exchange couplings, we have evaluated the specific heat in order to make an estimate of the transition temperature $T_N$. The results are shown in Fig. \ref{fig:specificheat} where we report the curves obtained for $U=0$ and $U=1\,$eV.
We see that for $U=1\,$eV we obtain a value of the critical temperature $T_{N}$=185 K that underestimates the experimental one, equal to $T_{N}\sim$ 265 K. We also tried with $U=2\,$eV but this choice led to no improvement of the result. Therefore, we conclude that the more reliable set of couplings is the one obtained choosing $U=1\,$eV, reported in Table \ref{tabJ}.
Although the calculated value of $T_{N}$ underestimates the experimental value found in CrAs, it is nonetheless close to that reported for CrAs$_{0.5}$Sb$_{0.5}$ \cite{Wang16}. 
We can thus say that this result, together with the one obtained for the propagation vector $\bm Q$, 
suggests that our approach allows to  
describe qualitatively the properties of this class of materials.

\section{Conclusions}
We have evaluated the magnetic exchanges of the pressure-induced superconductor chromium arsenide CrAs via {\it ab initio} LDA+U calculations, then mapping the energy of the magnetic configuration onto a Heisenberg Hamiltonian. 
The best agreement with the experimental results is obtained for a Coulomb repulsion $U=1\,$eV, this providing evidence that the system is a moderately correlated compound. This finding is consistent with the fact that the band structure is well reproduced by density functional theory calculations and that the value of the magnetic moment obtained for small $U$ agrees with the experimental one \cite{Autieri17,Autieri17b}. 
We have analyzed the double helix structure and in particular we have determined the temperature dependence of the specific heat by using Monte-Carlo simulations. 
The theoretical value of the critical temperature $T_{N}$ was found to be equal to 185 K, which underestimates the experimental value of 265 K. This discrepancy was to some extent expected, considering that it is well known that the use of an Heisenberg-like model alone cannot describe an itinerant magnet in a fully satisfactory way. Nonetheless, we notice that our {\it ab initio} approach allows to qualitatively reproduce the magnetic exchange couplings and the double helix structure with a propagation vector $\bm Q=(0,0,0.456)$. 
We expect that a further improvement of our results can be obtained by including in the approach presented here the dynamical effects related to the behavior of the chromium 3$d$ electrons as well as the Dzyaloshinskii-Moriya interaction. In particular, the latter may prove to be relevant due to the role played by the spin-orbit coupling in the arsenic atoms located between two chromium ones.


\section{Acknowledgments}
The work was supported by the Foundation for Polish Science through the International Research Agendas program cofinanced by
the European Union within the Smart Growth Operational Programme.

\end{document}